\documentstyle[psfig,sprocl]{article}

\def\Journal#1#2#3#4{{#1} {\bf #2}, #3 (#4)}

\def\NPA{{\em Nucl. Phys.} A}
\def\PLB{{\em Phys. Lett.}  B}
\def\PRL{\em Phys. Rev. Lett.}
\def\PRC{{\em Phys. Rev.} C}

\def\ZPA{{\em Z. Phys.} A}

\def\be{\begin{equation}}
\def\ee{\end{equation}}
\def\bea{\begin{eqnarray}}
\def\eea{\end{eqnarray}}
\def\etal{{\it et al.}}

\renewcommand{\vec}[1]{\mbox{\boldmath $#1$}}

\begin{document}

\title{SURFACE DIFFUSENESS ANOMALY \\ IN HEAVY-ION FUSION POTENTIALS}

\author{K. HAGINO}

\address{Yukawa Institute for Theoretical Physics, Kyoto Unversity, 
Kyoto 606-8502, Japan \\
E-mail: hagino@yukawa.kyoto-u.ac.jp}

\author{M. DASGUPTA, I.I. GONTCHAR, D.J. HINDE, C.R. MORTON, J.O. NEWTON}

\address{
Department of Nuclear Physics, Research School of Physical Sciences
and Engineering, Australian National University, Canberra, ACT 0200, Australia}


\maketitle
\abstracts{ 
Recent high precision experimental data for heavy-ion fusion cross
sections at energies in the vicinity of the Coulomb barrier
systematically show that a strikingly large surface diffuseness
parameter for a Woods-Saxon potential is required in order to fit the
data. We discuss possible origins for this anomaly, including the
effects of dissipation and the sensitivity of fusion cross 
sections to the choice of inter-nuclear potential. Our study suggests that 
the frozen density approximation, which is often used in analyses of heavy-ion 
reactions, may have to be re-examined for heavy-ion fusion. }

\section{Introduction}

One of the challenging problems in heavy-ion physics is to reproduce 
simultaneously experimental data for several processes, such as
elastic and inelastic scattering, fusion, and particle transfers, 
within a unified framework. A coupled-channels
approach~\cite{HRK99,T88,R81} provides an ideal tool for
this purpose. The most important ingredient in the coupled-channels
calculations is the inter-nuclear potential. A very popular parametrization
for this is the Woods-Saxon form (WS), given by 
\begin{equation}
V_N(r)=-V_0/[1+\exp((r-R_0)/a)]\, .
\end{equation}
This potential has been used widely in analyses of heavy-ion
reactions and has enjoyed success. 

For scattering processes, it has been well accepted that
the surface diffuseness parameter $a$ is around 0.63 fm.~\cite{CW76,S01} 
In marked contrast, recent high precision experimental data for
fusion cross sections, which were measured~\cite{DHRS98,L95,M99} 
with the aim of extracting
experimental fusion barrier distributions~\cite{RSS91},  
suggest that a much larger diffuseness parameter,
ranging between 0.8 and 1.4 fm, is needed in order to fit the data. 
This is not only for a particular system, but seems to be a
general trend.~\cite{L95,M99,M01,N01,SC01}

\begin{figure}[t]
\begin{center}
\leavevmode
\parbox{0.8\textwidth}
{\psfig{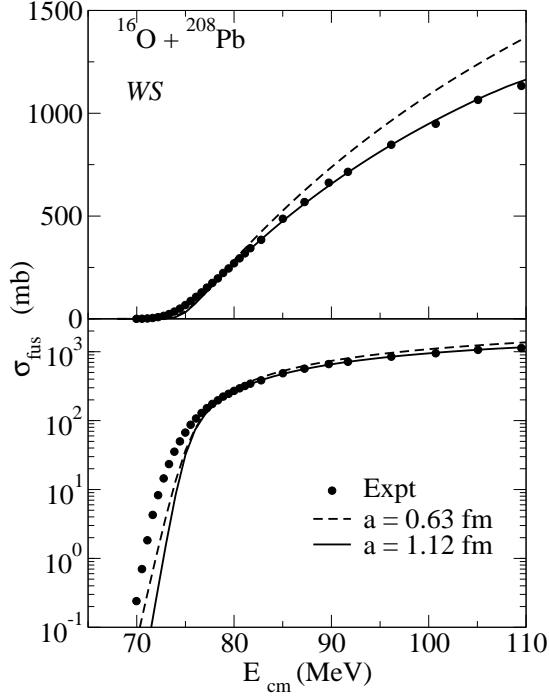}}
\end{center}
\protect\caption{
Comparison of potential model with the experimental data for 
the $^{16}$O + $^{208}$Pb fusion reaction. The upper panel is in the
linear scale, while the lower panel in the logarithmic scale. The
solid and the dashed lines are obtained by setting the surface 
diffuseness parameter
of Woods-Saxon potential to be $a$=1.12 and $a$=0.63 fm, respectively. 
The experimental data are taken from Ref. 8. }
\end{figure}

We illustrate this problem in Fig. 1 by comparing experimental data 
for the $^{16}$O + $^{208}$Pb fusion reaction~\cite{M99} with 
various calculations with the WS potential model, which 
includes the incoming wave boundary condition (see Sec. 2) but no channel 
couplings. 
The problem is twofold: at energies well above the barrier, 
where the fusion cross sections $\sigma_{fus}$ is relatively
insensitive to the couplings, a WS potential with a diffuseness 
$a$ =0.63 fm significantly 
overestimates fusion cross sections (dashed line). 
Changing the $V_0$ and
$R_0$ parameters is not helpful,
since it merely leads to an energy shift in the calculated fusion 
cross sections 
without significantly changing the energy dependence. 
On the other hand, a potential with $a$=1.12 fm (full line) fits the
data well. Below the barrier, the slope of the exponentially 
falling cross section is given well by the $a$=1.12 fm calculation 
but is not steep enough in that of $a$=0.63 fm. 
A better fit to this region would be given by including channel
couplings but only if the slope were correct. Thus both above and
below the barrier the much larger diffuseness parameter $a$=1.12 fm 
is required to match experimental data. 

The discrepancy may have a number of causes that include i) channel coupling
effects which were not included in the calculations, ii) the deviation
of the nuclear
potential from the Woods-Saxon parametrization, 
and iii) friction effects at high angular momenta. 
Of course, a potential with such a large diffuseness parameter should
be regarded as an effective one, which is appropriate only for
describing the particular channel. But, it is important to
understand the reasons for the large discrepancy in the diffuseness
parameters 
extracted from scattering and from fusion, in order to improve our
understanding of the fusion process and to allow reliable
predictions of subbarrier fusion cross section. 

Among the three possibilities above, 
the first point was discussed by Esbensen and Back.~\cite{EB96} 
They showed that the inclusion of collective excitations in both
the projectile and the target 
for the $^{16}$O + $^{144}$Sm reaction in the coupled-channels
calculations significantly improves the fit, leading to a surface
diffuseness of $a$=0.64 fm. However, the effects of channel coupling
on the surface diffuseness parameter of the inter-nuclear potential
were not shown to be significant for other systems, e.g. 
$^{16}$O + $^{148}$Sm reaction, as was discussed in Ref. 14. 

In this contribution, we study the other two possible explanations of the
surface diffuseness problem. In order to assess  
the sensitivity of fusion cross sections to the parametrization of
the inter-nuclear potential, and to discuss the effects of 
deviations from the Woods-Saxon shape, 
we will calculate in Sec. 2 fusion cross sections using a
double-folding potential~\cite{SL79,BS97,KS00}, 
as well as an error function.~\cite{N01,EWB83}
In Sec. 3, we will 
discuss the effects of friction on heavy-ion fusion
reactions using the critical distance model of Glas and 
Mosel~\cite{GM74} as well as the surface friction model.~\cite{F84} 
The summary is given in Sec. 4. 

\section{Choice of inter-nuclear potential}

In this section, we use the barrier penetration model to study the 
sensitivity of fusion cross sections to the choice of inter-nuclear 
potential. 
Instead of introducing an imaginary part to the inter-nuclear 
potential, in the calculations shown below, we use the so-called 
incoming wave boundary condition.~\cite{HRK99,LP84}
With this boundary condition, it is assumed that there is only an
incoming wave inside the Coulomb barrier due to a strong
absorption, which should be reasonable for heavy-ion reactions. 
In this model,
fusion cross sections are identified with the absorption cross
section. 

\subsection{Double-folding potential}

A double-folding potential~\cite{SL79,BS97,KS00} is constructed by 
folding an effective 
nucleon-nucleon interaction with the target and the projectile
densities: 
\begin{equation}
V_N(\vec{r})=\int\,d\vec{r}_1d\vec{r}_2\,v(\vec{r}_2-\vec{r}_1-\vec{r})
\rho_1(\vec{r}_1)\rho_2(\vec{r}_2)\, .
\end{equation}
This potential is intimately related to the resonating 
group method (RGM)~\cite{H91} and has been widely used in microscopic studies 
of heavy-ion reactions. 

\begin{figure}[t]
\begin{center}
\leavevmode
\parbox{0.8\textwidth}
{\psfig{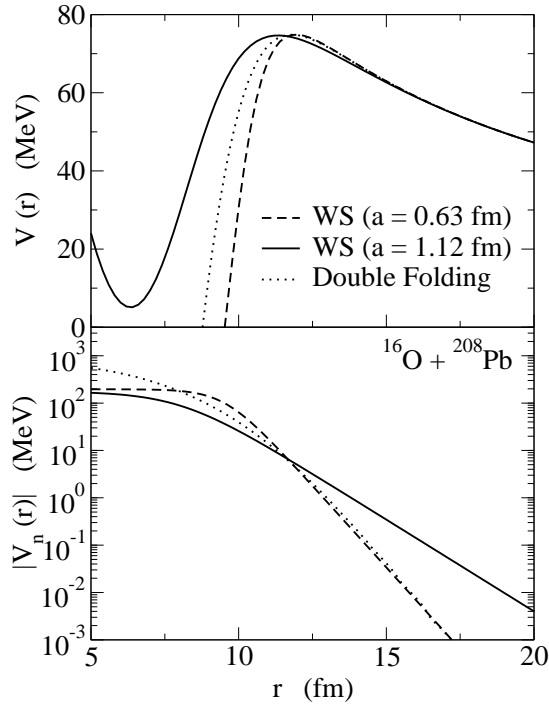}}
\end{center}
\protect\caption{
Comparison between the double-folding potential 
and Woods-Saxon potentials for the $^{16}$O + $^{208}$Pb reaction. 
The upper panel shows the total potential for $s$-wave scattering,
while the lower panel shows only the nuclear potential. }
\end{figure}

Figure 2 compares the double-folding potential 
for the $^{16}$O + $^{208}$Pb reaction 
with Woods-Saxon 
potentials for two different surface diffuseness parameters. 
For the nucleon-nucleon interaction $v$, 
we use
the density-dependent Michigan three-range Yukawa (DDM3Y1-Paris) 
interaction~\cite{KO95}, together with the zero-range approximation 
for the exchange contribution (see Ref. 16 for the parameters). 
We use a Fermi function for the $^{208}$Pb density as is given in 
Ref. 24, 
while the same density is used for $^{16}$O 
as that in Ref. 5. 
The double-folding potential roughly
follows the Woods-Saxon potential with $a$=0.63 fm, especially at
large distances, as would have been expected. 
It is in general much deeper than the Woods-Saxon
potential, and accordingly some deviation 
from the Woods-Saxon potential is seen in fig.2 at smaller distances. 
Figure 3 compares fusion cross sections for the double-folding
potential with those for the Woods-Saxon potentials. One can see that the
double-folding potential leads to similar fusion cross sections to a
Woods-Saxon potential with $a$=0.63 fm. The double-folding potential alone,
therefore, does not resolve the large diffuseness problem found
in heavy-ion subbarrier fusion reactions. 

\begin{figure}[t]
\begin{center}
\leavevmode
\parbox{0.8\textwidth}
{\psfig{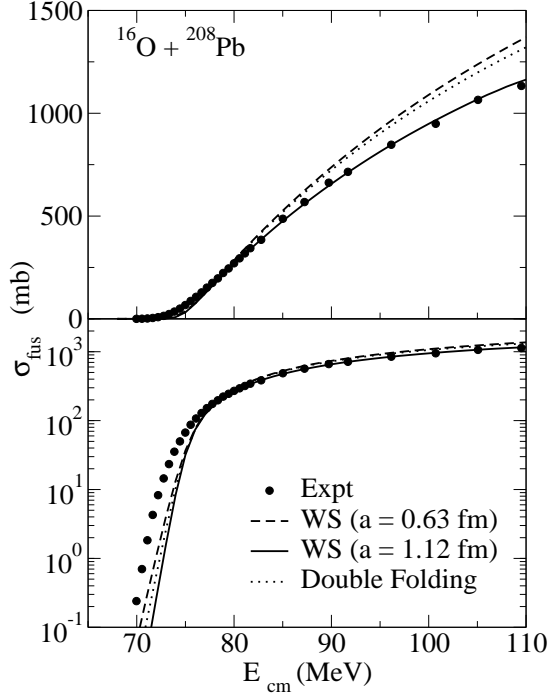}}
\end{center}
\protect\caption{
Fusion cross sections obtained with the double-folding potential and
their comparison with Woods-Saxon potential 
for the $^{16}$O + $^{208}$Pb reaction. }
\end{figure}

\begin{figure}[htb]
\begin{center}
\leavevmode
\parbox{0.8\textwidth}
{\psfig{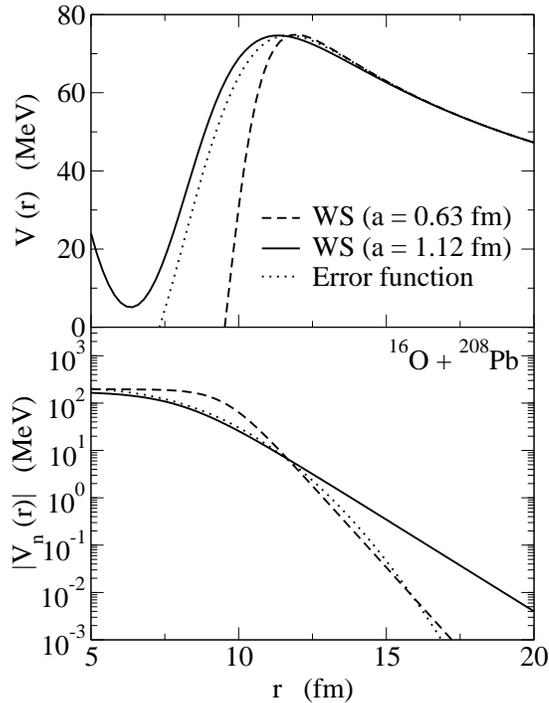}}
\end{center}
\protect\caption{
Same as fig. 2, but for an error function potential. }
\end{figure}

\subsection{Complementary Error Function Potential}

We next consider the error function potential defined by 
\begin{equation}
V_N(r)=-\frac{V_0}{2}\,{\rm erfc}((r-R_0)/a), 
\end{equation}
where the complementary error function is defined as 
\begin{equation}
{\rm erfc}(x)=1-\frac{2}{\sqrt{\pi}}\int^x_0dt\,e^{-t^2}.
\end{equation}
It has been pointed out in Ref. 11 
that this potential falls
more rapidly than the Woods-Saxon potential with increasing values of
$r$ while the curvature of the total potential remains similar. 
This is demonstrated in Fig. 4, where the error function potential
with $V_0$=220 MeV, $r_0$=0.92 fm, and $a$=2.9 fm is compared with the
Woods-Saxon potential (note that the $a$ parameter in the error
function potential (3) does not have a unique correspondence to the
surface diffuseness parameter in the Woods-Saxon potential). 
At distances larger than $r$=10.5 fm, this potential is very close to
the Woods-Saxon potential with $a$=0.63 fm, while at smaller distances
it is similar to the Woods-Saxon potential with $a$=1.12
fm. As a consequence of the latter, the barrier curvature is nearly the
same as that for the Woods-Saxon potential with $a$=1.12 fm, as can be
seen in the upper panel of fig. 4. 

\begin{figure}[t]
\begin{center}
\leavevmode
\parbox{0.8\textwidth}
{\psfig{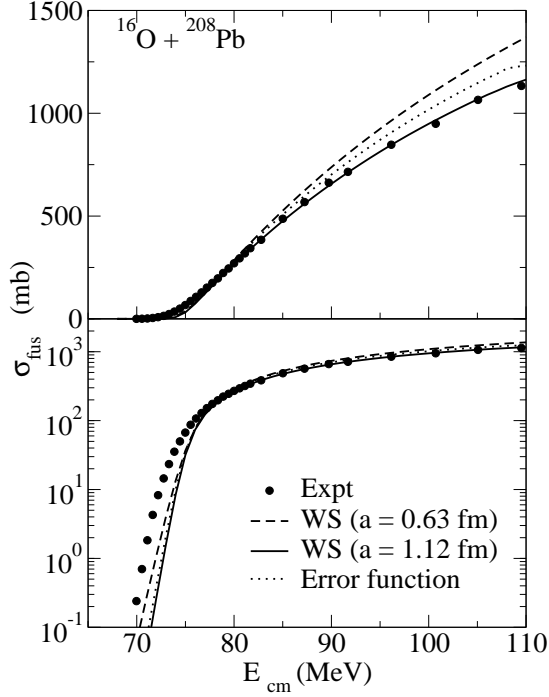}}
\end{center}
\protect\caption{
Same as fig. 3, but for an error function potential. }
\end{figure}

Fusion cross sections obtained with the error function potential are
shown in fig. 5. It is interesting to notice that this potential gives 
fusion cross sections close to the Woods-Saxon potential with
$a$=1.12 fm, thus providing an improved energy dependence of fusion
cross sections. As is shown in fig. 4, this potential is similar to
the Woods-Saxon potential with $a$=0.63 fm at large distances, and
therefore, would lead to similar scattering cross sections. 
The error function thus provides a promising
potential which may describe the scattering and fusion
processes simultaneously.~\cite{N01}

The main difference between the error function and the double folding
potentials is that the former falls off as a function of 
inter-nuclear distance in a
Gaussian way while the latter falls exponentially. In the folding
procedure, it is implicitly assumed that the reaction takes place much
faster than the dynamical density evolution. In this approximation,
the projectile as well as
the target densities remain unchanged even inside the
Coulomb barrier and this leads to 
an exponential tail of the inter-nuclear potential~\cite{AW79} 
as a consequence of 
the fact that the nuclear density behaves exponentially at the surface. 
This approximation is referred to as 
the frozen density approximation, and 
has been well tested in scattering processes.~\cite{NTG75}  
However, if the error function potential reflected the  
\lq\lq true \rq\rq shape 
of the inter-nuclear 
potential, its Gaussian tail would suggest a deviation from
the frozen density
approximation (i.e., an exponential tail) 
for subbarrier fusion reactions, and
effects of dynamical density evolution such as neck
formation~\cite{ABCD87} would have to be taken into account. 
This is certainly an interesting problem, but is beyond our intention 
in this contribution and we leave it for future study. 

\section{Effects of energy dissipation}

As an alternative origin for the large surface diffuseness problem, in
this section we discuss the effects of energy dissipation. 
A very simple and phenomenological way to mimic friction
effects is to use the critical distance model.~\cite{GM74,GGLT74} 
(See also Ref. 29). 
This model assumes that there is a certain distance $R_{c}$ which
has to be reached in order for fusion to occur. For each value of $E$, 
there is a corresponding critical angular momentum $l_c$ above which
$R_{c}$ is in the classically forbidden region as angular momentum 
is increased. Only partial waves
below $l_c$ then contribute to fusion cross sections. 
Approximating the Coulomb barrier by a parabola, and limiting the
partial wave summation up to the critical angular momentum, one obtains 
the expression for the fusion cross section in the Glas-Mosel model
~\cite{GM74}: 
\begin{equation}
\sigma_{fus}(E)=\frac{\hbar\Omega}{2E}R_b^2
\log\left\{    
\frac{1+e^{-2\pi(V_b-E)/\hbar\Omega}}
{1+e^{-2\pi(V_b-E+(E-V_c)R_c^2/R_b^2)/\hbar\Omega}}
\right\}, 
\end{equation}
where $\Omega, R_b$ and $V_b$ are the curvature of the Coulomb barrier, 
the barrier position, and the barrier height, respectively. 
$V_c$ is the value of the total potential at $r=R_c$. 
The critical angular momentum is given by 
\begin{equation}
E=V_c+\frac{l_c(l_c+1)\hbar^2}{2\mu R_c^2}. 
\end{equation}
Note that the Wong formula~\cite{W73} for the 
fusion cross section can be recovered
by setting $R_c$ to be infinity, that is equivalent to not restricting the
partial wave summation in computing fusion cross sections. 

\begin{figure}[t]
\begin{center}
\leavevmode
\parbox{0.8\textwidth}
{\psfig{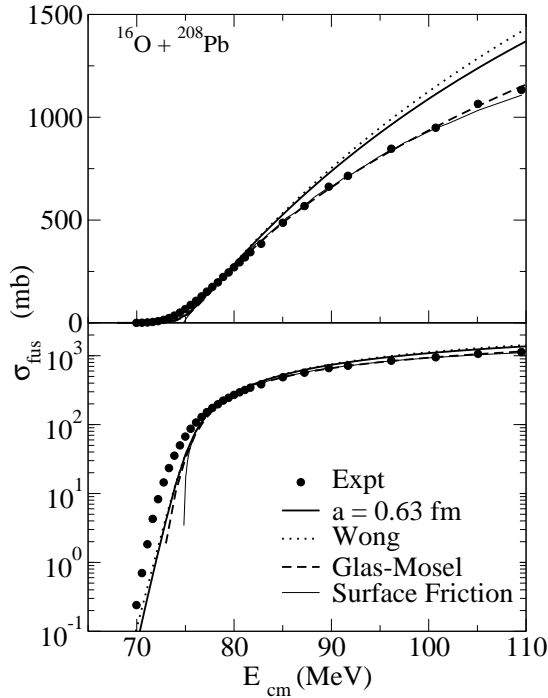}}
\end{center}
\protect\caption{
Fusion cross sections calculated with several models. The solid line is the
prediction of the barrier penetration model which uses the Woods-Saxon
potential with $a$=0.63 fm, while the dashed line is the result of the
Glas-Mosel model. The parabolic approximation to the barrier
penetration model is shown by the dotted line as a comparison. 
The thin solid line is the result of the surface friction model. }
\end{figure}

Figure 6 shows the results of analyses based on the Glas-Mosel model 
for the $^{16}$O + $^{208}$Pb reaction. 
We first examine the validity of the parabolic approximation used in the
Glas-Mosel model. The solid line is the result of the WS 
potential model with $a$=0.63 fm. If one
approximates this potential by a parabola, the barrier parameters are 
$R_b$=11.94 fm, $V_b$=74.87 MeV, and $\hbar\Omega$=5.00 MeV. The
dotted line shows the fusion cross sections obtained with the Wong
formula, which is based upon the parabolic approximation. 
Comparison between the solid and the dotted lines shows that
the parabolic approximation works reasonably well in the energy range shown
in the figure, although agreement is not perfect.  

We now vary $R_c$ and $V_c$ independently to fit the
experimental  data, whilst retaining the barrier parameters 
$R_b$, $V_b$, and $\hbar\Omega$. The dashed line is for $R_c$=10.39 fm
and $V_c$=72.4 MeV. 
Although the value of the critical distance $R_c$ is larger than the
systematics found in Ref. 19, 
this calculation reproduces the experimental data well at
energies above the barrier, and the slope of fusion cross sections 
below the barrier is
very similar to the prediction of the WS potential model 
with $a$=1.12 fm. Astonishingly, they are
almost indistinguishable if they are plotted in the same figure at this scale. 

A more physical but still classical approach to incorporating
dissipation is the surface friction model of Fr\"obrich.~\cite{F84} 
This model uses a single-folding potential, and friction coefficients 
which are propotional to the square of the first derivative of the nuclear
potential. With slight adjustment of the
inter-nuclear potential, and a changed tangential friction, quite good
reproduction of the data was obtained (see thin solid line). 

The surface-friction
model is a macroscopic phenomenological model, 
and does not consider any particular origin of dissipation. 
One of the main origins would be inelastic
excitations and/or particle transfer.~\cite{NT86} 
However, 
for the reaction $^{16}$O + $^{208}$Pb, 
the effects of channel
coupling are relatively weak due to the double closed shell structure of
the projectile and the target.~\cite{M99} 
Also it has been pointed out in Ref. 32 
that energy
dissipation due to particle
transfer is largely compensated with the change of potential due to the
polarization and thus the Coulomb barrier is reached almost without 
friction. Therefore neither collective
excitations nor particle transfers would lead to the large dissipation
effects, at least at the energies we are interested in. 
The only process which can be 
attributed to the origin of the dissipation would be a dynamical 
change in density of the reactants in the region where they overlap. 
This study again suggests a need to
go beyond the frozen density approximation. 

\section{Summary}

We studied possible origins of the recent finding that 
the experimental data for subbarrier fusion cross sections systematically
favour a much larger surface diffuseness parameter 
for a Woods-Saxon form of inter-nuclear 
potential than the commonly accepted value of $a \sim $0.63 fm. 
An obvious possibility would be that the inter-nuclear 
potential may deviate from
the Woods-Saxon potential. We
found that the double-folding potential roughly follows a Woods-Saxon
potential with $a$=0.63 fm, and thus does not account for the
large surface diffuseness parameter for subbarrier fusion. 
On the other hand, parameters could be adjusted for a complementary
error function potential so that it looked like a WS with a large
surface diffuseness at smaller distances while keeping 
$a \sim $0.63 fm at the surface. The error function, therefore, provides a 
promising potential which could simultaneously describe fusion and
scattering. It falls off in a Gaussian way with increasing 
inter-nuclear 
distance whereas a WS potential falls off exponentially. 
The exponential fall-off is a consequence of the frozen density
approximation. Hence, if the error function potential 
expresses the \lq\lq true \rq\rq
shape of the inter-nuclear potential, this may suggest a dynamical
evolution of density during fusion. 
As an alternative approach to the problem, we also examined the
dissipation effects using the phenomenological critical distance model
of Glas and Mosel as well as the surface friction model. 
The high energy data could be equally well reproduced both by these 
models and by using the WS potential with a large
diffuseness parameter. 
The fact that the data could be
explained both by changes in the potential, and the incorporation of
friction, indicates that it may not be a simple matter to disentangle these
two effects.
We argue that the dynamical density change during the fusion
process is one of the main origins of dissipation. This again suggests 
a deviation from the frozen density approximation, and thus a
transition from the sudden to the adiabatic potentials.~\cite{P72}
In this
connection, we would like to point out that in the adiabatic picture 
the inertia parameter may deviate from the reduced mass inside
the barrier, as was discussed long ago by Mosel.~\cite{M72}
Apparently, dynamical calculations are needed in order to draw a more
conclusive picture for the surface diffuseness problem. Such
calculations are now in progress, and we will report them in future 
publications.

\end{document}